\documentclass[11pt, a4paper]{article}
\usepackage{jheppub}
\usepackage{graphicx}
\usepackage{fancyvrb}
\usepackage{makeidx}
\usepackage{latexsym,amssymb,amsmath,graphicx, makeidx,slashed} 
\usepackage{epstopdf}

\newcommand{\beq}{\begin{equation}}
\DefineVerbatimEnvironment{code}{Verbatim}{fontsize=\small}
\newcommand{\eeq}{\end{equation}}
\newcommand{\bit}{\begin{itemize}}
\newcommand{\eit}{\end{itemize}}
\newcommand{\cL}{{\cal L}}
\newcommand{\cO}{{\cal O}}
\newcommand{\stp}{{\tilde t}}
\newcommand{\met}{{\slashed E_T}} 
\makeindex

\title{SUSY Stops at a Bump}
\author[a,b]{Christopher Brust,}
\author[c]{Andrey Katz,}
\author[a]{and Raman Sundrum}
\affiliation[a]{Maryland Center for Fundamental Physics, Department of Physics, University of Maryland, \\ College Park, MD 20742}
\affiliation[b]{Department of Physics and Astronomy, Johns Hopkins University, Baltimore, MD 21218}
\affiliation[c]{Center for the Fundamental Laws of Nature,  
Jefferson Physical Laboratory, Harvard University, \\ Cambridge, MA 02138}
\abstract{We discuss collider signatures of the ``natural supersymmetry'' scenario with baryon-number violating R-parity violation. We argue that this is one of the few remaining viable
incarnations of weak scale supersymmetry consistent with full electroweak naturalness.
We show that this intriguing and challenging scenario contains distinctive LHC signals, resonances of hard jets in conjunction with relatively soft leptons and missing energy,
which are easily overlooked by existing LHC searches.
 We propose novel strategies for distinguishing these signals above background, and 
estimate their potential reach at the 8~TeV LHC. 
We show that other multi-lepton signals of this scenario can be seen by
currently existing searches with increased statistics, but these opportunities are more spectrum-dependent.}
\preprint{UMD-PP-012-008}
\begin{document}
\maketitle

\section{Introduction}
The scenario of 
weak scale supersymmetry (SUSY) with
 light third generation superpartners (``natural SUSY'' or ``effective SUSY'') 
 was put forward many years ago~\cite{Dimopoulos:1995mi,Cohen:1996vb} as a way of ameliorating the SUSY flavor and CP problems, while maintaining electroweak (EW) naturalness. The central observation from the hierarchy problem viewpoint is that, in the Standard Model (SM), the radiative corrections which destabilize the Higgs potential are dominated by the heavy top quark. EW stability requires SUSY cancellations from stops of a few hundred GeV. 
However, from a purely bottom-up viewpoint it is attractive to consider the other superpartners to be substantially heavier, since they are primarily implicated in the SUSY flavor and CP problems. The theoretical challenge is to then identify high-energy models and mechanisms which, at least approximately, lead to this kind of split superspectrum (see~\cite{Sundrum:2009gv,Barbieri:2010pd,Redi:2010yv,Craig:2011yk,Csaki:2011xn,Csaki:2012fh,Craig:2012di} for a partial list). 

There has been a resurgence of interest in this scenario in the last year, mainly motivated by the null results thus far of LHC searches for physics beyond the SM.  In light of the current LHC bounds on R-parity conserving SUSY, it is difficult to envision any other viable version of SUSY which is consistent with full electroweak naturalness (that is, absence of EW fine-tuning).
By contrast, it has been recently shown that natural SUSY easily evades the most stringent LHC constraints with integrated luminosity $\cL \sim 1$ fb$^{-1}$~\cite{Kats:2011qh,Brust:2011tb,Papucci:2011wy,Essig:2011qg}. Later dedicated searches for R-parity conserving natural supersymmetry have appeared, better constraining particular spectra with~\cite{Atlas:blmet} or without~\cite{Aad:2011cw} light gluinos, however the natural parameter space is still quite open.

Natural SUSY has an even wider significance, in that it beautifully illustrates the general theme of how a ``top-partner'' can algebraically cancel destabilizing top-quark radiative corrections to the Higgs potential. This relates it to the theory and phenomenology of fermionic top-partners~\cite{Berger:2012ec},  appearing in non-supersymmetric Little Higgs (see for review~\cite{Schmaltz:2005ky,Perelstein:2005ka} )and Twin-Higgs models~\cite{Chacko:2005pe,Chacko:2005vw}.
 In this sense, light stop searches fit into the broader program of testing whether {\it any} top-partner is helping to stabilize the weak scale. The LHC is the first experiment in history that can test naturalness on such a broad front, in a relatively comprehensive and well-defined way.  {\it Either a discovery of such top-partners at the natural scale of a few hundred GeV, or even their exclusion to high confidence, would constitute a significant scientific finding.} In previous work~\cite{Brust:2011tb}, we have argued that for this grand and challenging experimental undertaking, one should free the natural SUSY setting from too much UV prejudice and anticipation, lest this lead to overlooking experimental opportunities {\it now}
and because UV considerations have not led to any sharp no-go ``theorem''. We have carved out a simple theoretical framework that facilitates this. 
The present paper will discuss important but widely overlooked signals 
that follow straightforwardly from this perspective, as well as the combination of  methods that can separate them from SM background. 

In particular, 
although most experimental searches for natural supersymmetry have concentrated on the R-parity conserving case, we consider here the case of (baryon-number violating) R-parity violation (RPV). The plausibility and attractiveness of this scenario was argued in Ref.~\cite{Brust:2011tb} from a number of viewpoints and considerations (also see~\cite{FileviezPerez:2011pt} and references therein for recent models of spontaneous RPV).
The spectrum of RPV and R-parity conserving natural SUSY can be quite similar. If the theory is completely natural we  expect both species of stops and at least the left-handed sbottom  with masses of order $\sim \nolinebreak 400$~GeV or lighter. 
Gluinos can be naturally twice as heavy
if they are Majorana fermions, and even heavier if they are (part of) Dirac states~\cite{Fox:2002bu}.  That is, we cannot guarantee the gluino to be experimentally accessible in the near future. However, if we are lucky and Majorana gluinos are light enough in RPV natural SUSY, they can produce spectacular signatures in same-sign dileptons~\cite{Allanach:2012vj}. Here, we assume more minimally that
 the stops and  sbottom mandated by naturalness are the only accessible colored superpartners. 

RPV is distinct from R-symmetry conservation because phenomenological viability does not require a neutral superpartner to be at the bottom of the SUSY spectrum, since  superpartners are allowed to decay into SM particles. Therefore EW gauginos can easily be heavier than the stops and the sbottom, having no significant impact on the phenomenology.

Non-minimal Higgs degrees of freedom are subtler.
Higgsinos are usually assumed to acquire mass from the same ``$\mu$-term'' that also contributes to Higgs scalar potential. Higgs naturalness then requires Higgsinos not much heavier than $\sim 200$ GeV. (However, see Ref.~\cite{Brust:2011tb} for a bottom-up description in which Higgsinos can be much heavier.)  We will show in Sec.~\ref{sec:theory} that light Higgsinos can remain relatively well-hidden in the RPV context. 
On the other hand,  extra Higgs {\it boson} degrees of freedom of SUSY can be heavier without compromising naturalness. We therefore only keep the SM Higgs scalar in our study of stop/sbottom phenomenology. Finally, there are by now stringent bounds on the SM Higgs mass, and even tentative hints of its presence at $\sim 125$ GeV. Theoretically accommodating the Higgs mass in {\it high-energy} models has been an increasing challenge ever since LEP2, but there are certainly interesting ideas for doing this. By contrast, a $125$ GeV Higgs mass is straightforwardly accommodated within our bottom-up natural SUSY framework, deferring the full UV description of physics lying outside experimental reach. 
In particular, we do not restrict stop/sbottom masses by their radiative contributions to the physical Higgs mass, since there may well be other contributions from unknown heavier sources.

Collider signatures of RPV SUSY are largely dictated by the detailed structure of RPV interactions, which  cannot be anarchical (for a review see~\cite{Barbier:2004ez}). Either baryon number or lepton number should be conserved to avoid prompt single proton decay. While lepton number violation (LNV) is interesting by itself and deserves more study, it has already meaningful constraints from the LHC, since it mostly leads to leptons and taus in the final states, which are relatively easy to spot. Baryon number violation (BNV) is experimentally more challenging than LNV, resulting in jetty final states and suffering from enormous QCD and $t\bar t$ backgrounds. 
We will focus on this challenging scenario of two stops and a sbottom at the bottom of a BNV SUSY spectrum. Not only does this fill a gap in SUSY searches, but it also shares several features with, and insights into, other top-partner searches. 
This spectrum was considered earlier in Ref.~\cite{Kilic:2011sr} in the context of the CDF ``$Wjj$ anomaly''~\cite{Aaltonen:2011mk}. Although the anomaly was later refuted by D0~\cite{Abazov:2011af}, as well as by a similar CMS search~\cite{CMS:wjj}, this paper was an important step in understanding of collider signatures of the minimal spectrum. Here, we will elaborate on several points briefly touched on in Ref~\cite{Kilic:2011sr},  broaden the motivations and scope, and detail new search strategies. 

A particular difference with Ref~\cite{Kilic:2011sr} is that we will assume that BNV is governed by small couplings and therefore we will neglect single-resonance production of superpartners, concentrating on pair-production. Such smaller couplings make the theory more straightforwardly safe from low-energy precision data. 
We will argue that it is theoretically very plausible that a stop is the lightest superpartner, which can decay into a pair of jets.
 Needless to say, by itself this is an extremely challenging signature, given relatively small production cross sections and absence of any ``interesting'' features in the event, e.g. leptons or missing transverse energy (MET). However,  one can take advantage of production of the heavier sbottom and stop which further cascade decay into the lightest stop, emitting $W, \ Z$ and/or higgs (on-  or off-shell) along the way. These events are more promising, because they can potentially contain leptons and MET. Nonetheless, existing cut-and-count searches are not optimized for signatures like this and generally overlook them. They do not take advantage of the most important qualities of these events: hard jets reconstructing a pair of {\it resonances}, in conjunction with leptons and/or MET which is relatively soft compared to the top quark background.

We will substantiate these claims, and use them to craft a search strategy for the most promising and robust of these cascades. We will show that the backgrounds are under control and $\sqrt{s} = 8$ TeV LHC can have a good reach for these events. We will also discuss other channels, which can be promising, but where the backgrounds are not easy to estimate with theoretical tools.  The 
alternative possibilities for the lightest superpartner, a sbottom or Higgsino, are also plausible but even more phenomenologically 
challenging, and we defer their consideration from this paper. 

The paper is organized as follows.
In the next section we review the BNV RPV natural SUSY scenario and reduce it to its most LHC-relevant features, thereby arriving at a useful ``simplified model''. In section~\ref{sec:signals} we study the various ``charged current'' ($W$) channels and relevant backgrounds, and roughly estimate which of these channels is viable. In section~\ref{sec:reach} we perform explicit simulations of signal and background in the most promising of these channels and discuss cuts (which are quite different from standard SUSY searches) in greater detail. In section~\ref{sec:neutral} we briefly discuss ``neutral current'' ($Z, h$) cascade decays between stops. 
While getting a substantial number of these events is only possible in a subset of stop/sbottom/higgsino spectra, they can be quite spectacular, and indeed 
the multi-lepton CMS search~\cite{Chatrchyan:2012ye} already has an appreciable sensitivity. They  remain an  exciting discovery channel for the future with more statistics.  Finally in section~\ref{sec:conclusions} we conclude.

\section{Reduction to Simplified Model}
\label{sec:theory}

{\Large \it Spectrum}
\vspace{3mm}
\nopagebreak

 \noindent As we argued in~\cite{Brust:2011tb}, the only superpartners robustly required by naturalness to lie under $500$ GeV are (in EW gauge basis) the $\tilde{t}_R, \tilde{q}_L \equiv (\tilde{t}_L, \tilde{b}_L)$ stops and sbottom, and $\tilde{H}_u, \tilde{H}_d$ higgsinos (if their mass arises
from a $\mu$ term). Along with a SM Higgs boson $h$, we shall consider these the only new particles substantially accessible to the $7-8$ TeV LHC with moderate luminosity.

Colored superpartners have strong production cross-sections, which suggests 
that we focus our searches on them. As can be seen in Table~\ref{tab:Xsec}, even these strong cross-sections peter out for squark masses above $500$ GeV, so that the natural regime for the spectrum is also our only hope for direct visibility.
We will argue that light higgsinos are typically a complication in these searches, either mild or major depending on the spectrum, but rarely do they present a spectacular new opportunity. For now we simply neglect them, but return at the end of this section to better justify this position.

The gauge-basis squark states are non-trivially related to the mass-eigenstates after EW symmetry breaking, due to two effects, 
the splitting of $\tilde{t}_L$ from $\tilde{b}_L$, and the mixing of $\tilde{t}_L$ and $\tilde{t}_R$. The first of these effects is given by the sum of $F$ and $D$ term potentials, 
\beq
 m_{\tilde t_L} - m_{\tilde b_L} \approx  \frac{m_t^2 }{ m_{\tilde b_L} + m_{\tilde t_L}} - \frac{m_W^2 \sin^2 \beta}{ m_{\tilde b_L} + m_{\tilde t_L}}~. 
\eeq 
The $\tilde{t}_L-\tilde{t}_R$ mixing arises from a possible SUSY-breaking $A$-term, 
$A \tilde{t}^c_R h_u \tilde{q}_L$. This results in mass eigenstates 
$\tilde{t}_1$ and $\tilde{t}_2$ (with $m_{\tilde{t}_1} < m_{\tilde{t}_2}$ by convention) related to the gauge-eigenstates by an angle $\theta_{\tilde{t}}$. Combining these effects, the sbottom is either  the middle or the lightest of our squarks.
We focus on the former case, with a 
 spectrum $\tilde t_2-\tilde b - \tilde t_1$. Note also, that even though the mass splitting between $\tilde{b}$ and $\tilde{t}_1$ is essentially a free parameter,  it cannot  be too big if we play by two rules: (a) we keep the lightest squark heavier than the top quark, which would otherwise change the phenomenological possibilities, and (b) keep the spectrum natural and production cross-sections for $\tilde{b}$ appreciable by not making it too heavy.

\vspace{5mm}
\noindent {\Large \it Couplings}
\nopagebreak
\vspace{3mm}
\nopagebreak

\noindent The central novel interaction being considered is the BNV RPV coupling,
\begin{equation} 
{\cal L} \supset \frac{\lambda''_{3IJ}}{2} \tilde{t}_R^c d_R^{cI} d_R^{cJ} + {\rm h.c.},
\end{equation}
 enabling stop decay to SM quarks.
The exact flavor structure of these couplings is constrained by a variety of low-energy flavor and precision tests. From the viewpoint of LHC visibility the central issue is whether the quarks in the dominant BNV couplings carry heavy flavor ($b_R$ quarks) or not.  We have argued in Ref.~\cite{Brust:2011tb} that in either case, low-energy constraints can be satisfied within quite plausible UV flavor paradigms. Relatedly, in the case where all three generations of squarks are present, suitable flavor paradigms have also been studied by Refs.~\cite{Csaki:2011ge,KerenZur:2012fr}. We shall therefore consider two cases for stop decay to quarks via BNV: (a) one $b$ quark and one light quark, and (b) 
two light quarks.

We will always consider the generic possibility that there is at least modest non-vanishing $A$-induced mixing, such that both $\tilde{t}_{1,2}$ inherit BNV couplings to quarks via their $\tilde{t}_R$ component. Similarly, they both have
 weak couplings to the sbottom via their $\tilde{t}_L$ component. Furthermore,  mixing also leads to $\tilde{t}_1^c  \tilde{t}_2 Z$ and $\tilde{t}_1^c  \tilde{t}_2 h$ couplings, determined by the A-terms and mixing angles.  
We assume small BNV couplings, $\lambda''_{3IJ} \ll 1$, so that in general squarks will only decay through such interactions if decay by $W, Z$ or $h$ emission is kinematically suppressed, as for example is obviously the case for $\tilde{t}_1$ (which, recall, we are considering  as the lightest superpartner).

Since $\lambda''_{3IJ} \ll 1$ is the most straightforward way to comply with low-energy constraints,\footnote{Although one can easily satisfy low-energy constraints with $\lambda''_{3IJ} \ll 1$, there is still a concern that BNV can wash out the cosmological baryon asymmetry, if this is generated in the early Universe. One can try to turn this into a mechanism for actually generating the baryon  asymmetry below the EW scale~\cite{Dimopoulos:1987rk,Cline:1990bw,Scherrer:1991yu,Adhikari:1996mc}. A new robust approach will appear in~\cite{YanouRaman}.} it is important to ask how small these couplings can be without resulting in
displaced vertices at the LHC from a long $\tilde{t}_1$ lifetime.  
In order to not have a displaced vertex, we need $c/\Gamma$ to be less than about $1$ mm. The expression for the distance traveled before decay by a pure $\tilde{t}_R$ particle (ignoring mixing) is
\beq L 
\sim (1 {\mathrm ~ mm })\left(\frac{300\mathrm{~ GeV}}{m_{\tilde{t}_R}}\right)\left(\frac{(2.5\cdot 10^{-7})^2}{\sum \lambda_{3IJ}''^2}\right) 
\eeq
Thus, for $300$ GeV squarks, we need $\lambda''$ to be roughly bigger than $2.5\cdot 10^{-7}$. If the BNV couplings are smaller than this bound, we will have events with {\it jets} emerging from displaced vertices, which can further help discriminate against background. We will tackle the more challenging case in this paper, by assuming that the BNV couplings are strong enough that $\tilde{t}_1$ decays are prompt. 

\vspace{5mm}
\noindent {\Large \it Higgsinos}
\nopagebreak
\vspace{3mm}
\nopagebreak

\noindent We are now in a position to understand how higgsinos might affect the LHC physics. 
We have discussed how the spectrum of squarks can  be produced and then cascade decay by EW boson emission, with a final prompt BNV decay to quarks. 
Higgsinos of comparable mass to the squarks allow these steps to potentially be bypassed, by opening up alternative squark decays to higgsinos. 

The simplest case would be if the higgsinos were even a little heavier than the stops and sbottom. Since direct EW production has substantially lower cross-section, such higgsinos would be phenomenologically irrelevant. But if the higgsinos are lighter than the heaviest stop, then  $\tilde t_2$ decays via EW emission or BNV can be substantially degraded by decay to $\tilde{H}^{+} b$. (The alternate decay to  $\tilde{H}^{0} t$ is likely to be phase-space suppressed.) In turn, $\tilde{H}^{+}$ will decay (via $\tilde{t}_1$ and BNV) to three jets. In this way, the higgsinos will degrade events with leptons from (possibly off-shell) 
$W,Z$, and add events with extra $b$ jets. This is the basic complication we alluded to earlier: higgsinos can force us to look in multi-jet events, without spectacularly high $p_T$, with resonances obscured by combinatorial background, and with only the handle of several $b$ jets. 

But fortunately, the higgsinos can easily {\it not} degrade sbottom decays even if they happen to be lighter than the sbottom but heavier than the lightest stop, because 
the only sbottom decay to higgsinos (for small bottom Yukawa coupling) is to $\tilde{H}^{-} t$, which is likely highly phase-space suppressed. Unfortunately if the Higgsinos are at the bottom of the spectrum they will be produced in abundance in $\stp_1 \to b \tilde H^+$ decays. This decay mode does not affect the $W$ production but complicates the resonance reconstruction from the jets.    
This is one of the reasons that we focus on sbottom charged current decays in this paper: the phenomenology of  
sbottom $\rightarrow W^{(*)} \tilde{t}_1$ is largely ``immune'' to higgsinos if they are \emph{not} the lightest SUSY particles.

We proceed by dropping higgsinos from the discussion as part of arriving at our simplified model of stops and sbottoms in Sections~\ref{sec:signals} and~\ref{sec:reach}. As discussed above, this will only modestly affect our central channel of sbottom production and cascade decay, and is the best case for the other channels (but dependent on the spectrum). We will return however to the possibility of Higgsinos in the spectrum in Sec.~\ref{sec:neutral} since higgsinos can easily dramatically reduce the contributions of the neutral current decay $\tilde t_2 \to Z^{(*)} \stp_1$.  In particular the higgsinos can suppress a  yield of neutral current decays in an otherwise spectacular multilepton channel.

\section{Signals and strategy of search}
\label{sec:signals}
One finds the highest production cross sections for the lightest particles, which would imply in our case a search for  pair-production of the lightest stop with subsequent decays into four jets. However a search for resonances in 4-jet events is very challenging at the LHC~\cite{Aad:2011yh}, because it has to deal with a big uncertain QCD background, and even the multijet trigger is probably not 100\% efficient in this case.\footnote{An analogous search for resonances in 6-jet events~\cite{CMS:6jets} has a better reach, but it is relevant only for light gluinos in an RPV spectrum.}  Therefore it is fruitful to concentrate on longer cascades, which involve Higgs or EW boson emission (either on- or off-shell). This means we consider production of heavier states, $\tilde b, \ \tilde t_2$ and subsequent transitions 
\beq\label{eq:processes}
\tilde b \to W^{(*)} \tilde t_1, \ \ \tilde t_2 \to Z^{(*)} \tilde t_1, \ \ \tilde t_2 \to h^{(*)} \tilde t_1~.
\eeq 

While the first process in Eq.~\eqref{eq:processes} is fairly robust, the branching ratio of the two other  processes is model-dependent. The relative rate between the second and the third process in Eq.~\eqref{eq:processes} is determined by the couplings of the stops to $Z$ and $h$ and by phase space effects. While \emph{the neutral current decays can have spectacular multi-lepton signature} (see Sec.~\ref{sec:neutral} for a detailed discussion) it might also happen that \emph{the second stop's mass is between 400 and 500 GeV, rendering the production  cross-section tiny} (see Table~\ref{tab:Xsec}). Moreover most of the spectacular signatures come from the decays into $Z$ rather than the Higgs, though it would be very nice to eventually observe the Higgs in these new physics processes. We might however find ourselves in the situation that the mixing angle between the stops, $\theta_{\tilde{t}}$ is large and Higgs transitions are preferred.  Therefore, it is fair to say that the charged current transition is the robust and  spectacular channel at the 7-8 TeV LHC and we will give it most of our attention.

Before we continue with a detailed analysis of the cascade decays, we note that the table~\ref{tab:Xsec} pair-production cross sections for stops  were calculated at the NLO with {\tt Prospino~2.1}~\cite{Beenakker:1996ed}.  Sbottom production cross sections are usually slightly bigger due to electroweak corrections. These numbers will further help us in our numerical estimates. 

\begin{table}[t]
\centering  
\begin{tabular}[t]{|c|c|c|c|c|c|}
\hline 
{\small $m = 230 $ GeV} &  {\small $m = 250 $ GeV} &  {\small $m = 270 $ GeV} & {\small $m = 300 $ GeV} & {\small $m = 400$ GeV} & {\small $m = 500$ GeV}\\
\hline  
 8.3 pb & 5.3 pb  & 3.4 pb & 1.9 pb  & 0.34 pb & 0.08 pb \\
\hline 
\end{tabular}
\caption{Pair production cross sections for stop at NLO for center of mass energy $\sqrt s = 8$ TeV. Sbottom production cross-sections are very similar, slightly bigger though due to electro-weak effects. }
\label{tab:Xsec}
\end{table}
Charged-current $\tilde b \to W \tilde t $ transitions with a subsequent RPV decay of the stop into two jets were first addressed in~\cite{Kilic:2011sr} in the case of resonant production of sbottom. (Pair-production was also briefly considered, but was not the primary focus.) Although resonant production is not categorically excluded by the bounds on RPV, it requires strong enough BNV to raise FCNC and $n - \bar n$ oscillation concerns. However we focus  on pair production with RPV decay mediated by couplings which can be much smaller than one, and therefore safer from low-energy tests.

\begin{figure}[t]
\centering
\includegraphics[width=0.45\textwidth]{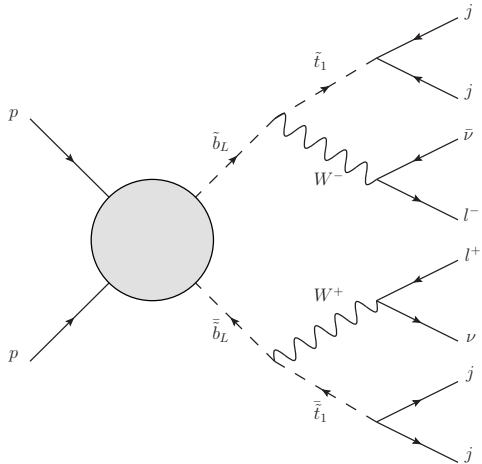}
\caption{The sbottoms are pair-produced and undergo charged-current decay. When both $W$s (either on- or off-shell) decay leptonically, they leave a spectacular signature of two leptons + jets, which reconstruct two equal-mass resonances. We analyze this signal in Sec.~\ref{sec:signals} and~\ref{sec:reach}.}
\label{fig:feyndiag}
\end{figure}

In this case the most spectacular signature shows up when both $W$s decay leptonically, leading to a signature $l^+ l^- jjjj + \met$, where the jets reconstruct two resonances with equal masses (see diagram in Fig~\ref{fig:feyndiag}). 
What should be our search strategy for these events? Performing cut-and-count search on events which reconstruct resonances is probably not ideal. However we can try to reconstruct resonances with the following steps:
\begin{itemize}
 \item Find events with 2 isolated leptons and moderate $\met $ (the latter should be non-zero to remove the background from DY dilepton production).
 \item Cluster the jets with sufficiently big radius (otherwise there is a danger that we lose the hadronic activity which reconstructs the resonance and thereby get edges instead of peaks).
 \item If the event contains 4 jets (or more), try all possible pairings between the jets, and pick up the combination which minimizes the difference between the reconstructed invariant masses. Discard the event if the minimal possible mass difference is too big. This step is essentially identical to the standard multi-jet resonances search~\cite{Kilic:2008ub}.
\end{itemize}

Unfortunately our events with 2 leptons, MET and multijets have an appreciable background, on top of which we are looking for our bumps. This background is heavily dominated by dileptonic $t \bar t$ (including $l \tau_l $ decay modes). One can show that with an adequate choice of cuts all other backgrounds ($Z \to \tau_l \tau_l~+$~jets, DY dileptonic production with jets, $WW~+$~jets) are highly subdominant to $t \bar t$, and we will discuss it in more detail in the next section. Production cross section for dileptonic $t \bar t$ exceeds our signal by two orders of magnitude, and  even though the extra jets in these events do not come from  resonances, reconstructing ``by accident'' two pairs of jets with similar invariant masses is common.   
The above mentioned  steps, plus standard cuts for the overall hardness of the event, 
are still not enough in order to see clear bumps on top of this continuous $t \bar t$ background after $\sqrt{s} = 8$ TeV run.  
We therefore use other, less standard discriminators to distinguish the signal from the background. 

\begin{figure}[t]
\centering
\includegraphics[width=0.49\textwidth]{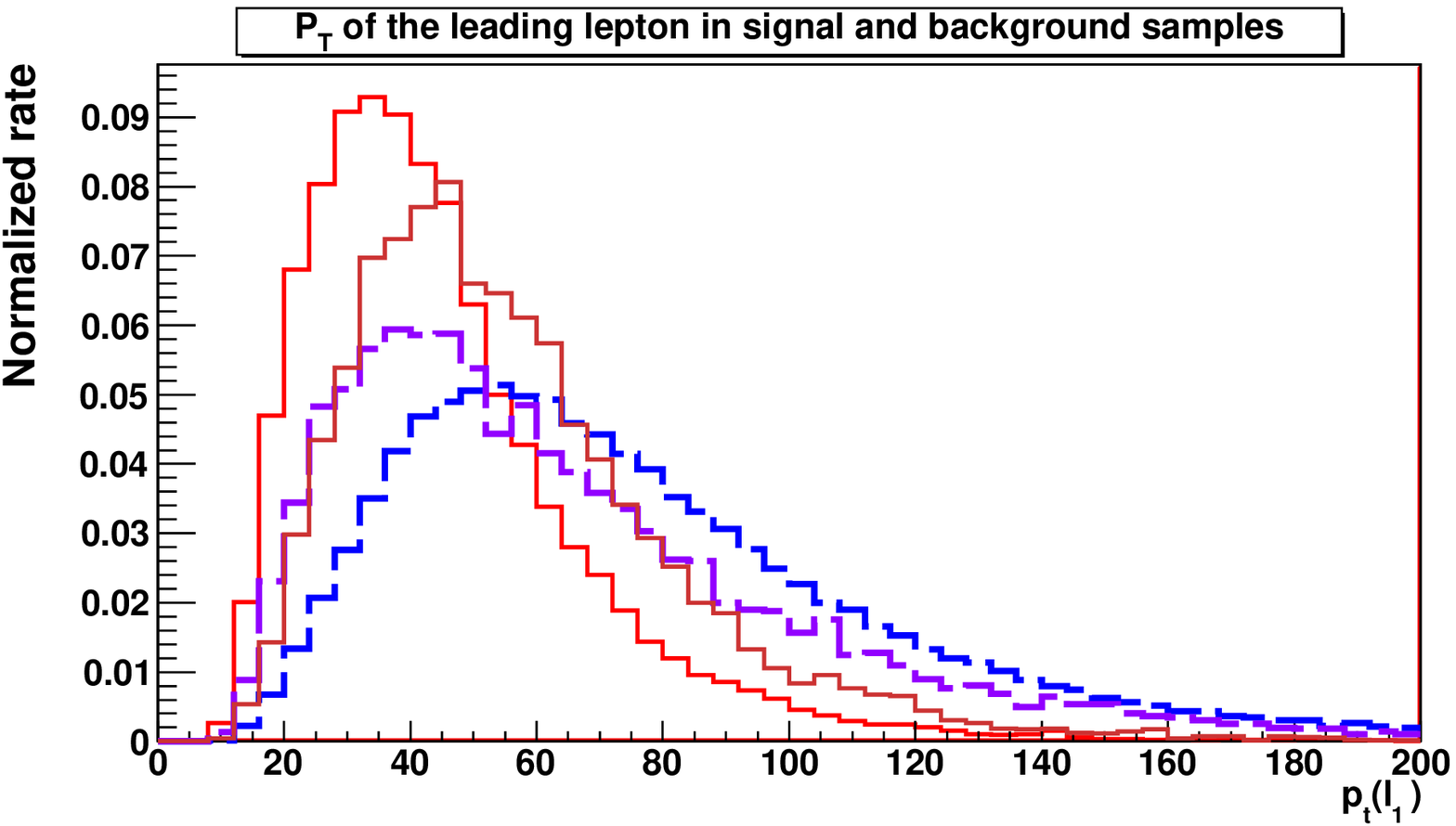}
\includegraphics[width=0.49\textwidth]{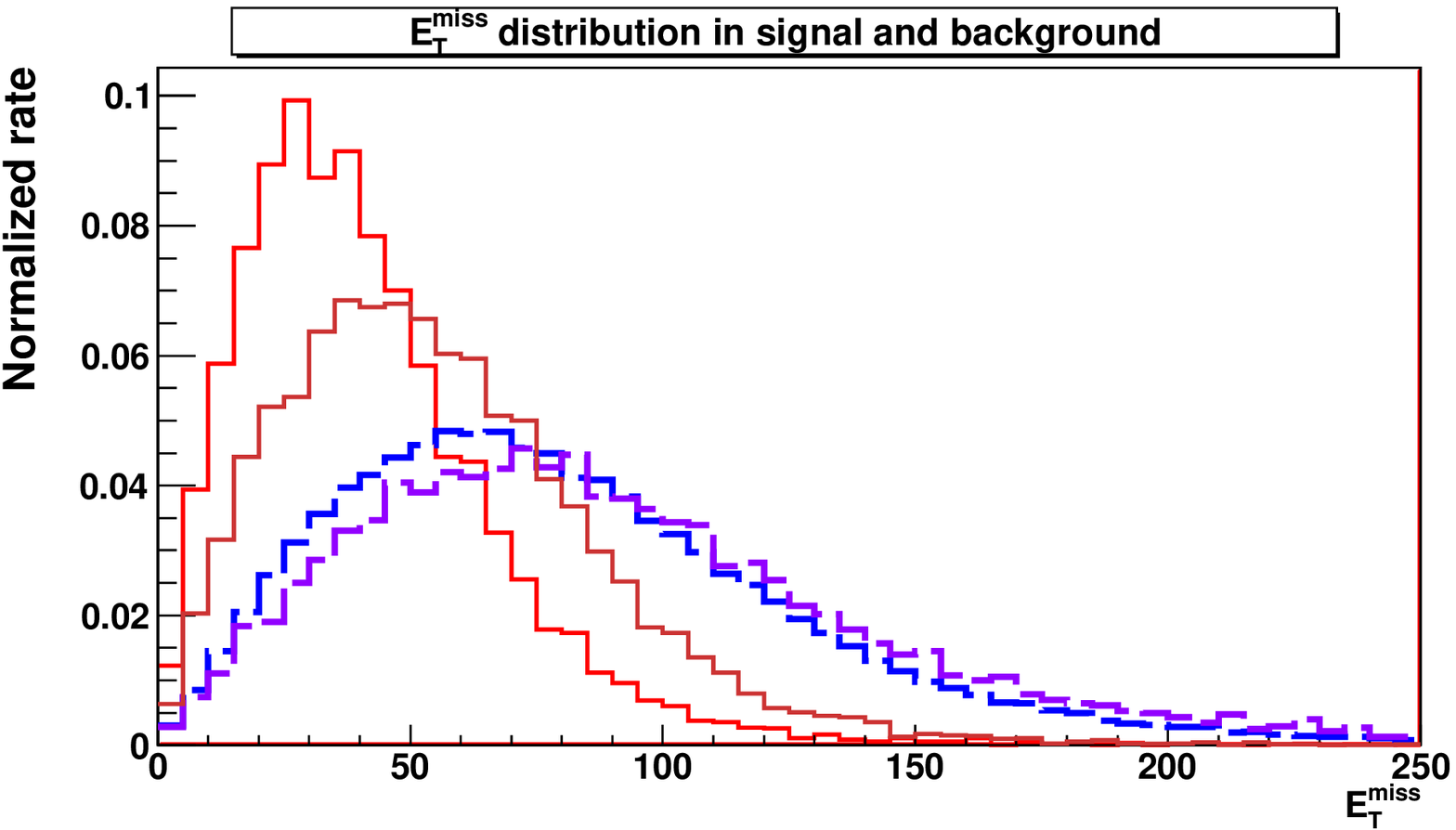}
\caption{Leading lepton $p_T$ and $\met$ distribution in signal and background events. Light red and dark red curve stand for the signal spectra 1 and 2 (see Table~\ref{tab:benchWdecay}). The blue line represents the distribution in dileptonic $t \bar t$ background and the violet line represents the $l \tau_l$ background (which we simulate separately since it has slightly different kinematics). See Sec.~\ref{sec:reach} for details of simulations.}
\label{fig:ptl1_met_absolute}
\end{figure}

There are two additional important features which distinguish our signal from the background. Usually in a dileptonic $t \bar t$ event, hardness of the entire event correlates with the hardness of the leptons and the $\met$. This happens because the $W$ is often boosted in the rest frame of the decaying top. However it is not the case in the signal. As we have explained in Sec.~\ref{sec:theory}, naturalness and visibility  motivate mild splittings between the stop and the sbottom, usually  so small that they  do not allow emission of the \emph{on-shell} $W$. Even if emission of the on-shell $W$ is allowed it typically has little boost in the rest frame of the decaying sbottom. This results in relatively small $p_T(l)$  and $\met$ even if the event overall is very hard. We demonstrate the distribution of $\met$ and the transverse momentum of the leading lepton in signal and background events on Fig.~\ref{fig:ptl1_met_absolute}. This immediately suggest that just cutting on the tail of high $\met $ and high $p_T(l_1)$ should be a decent discriminator between the signal and the background. We checked it explicitly and it indeed removes a fair portion of the background. We will use   
a refined version of this discriminator below. 

\begin{figure}[t]
\centering
\includegraphics[width=0.49\textwidth]{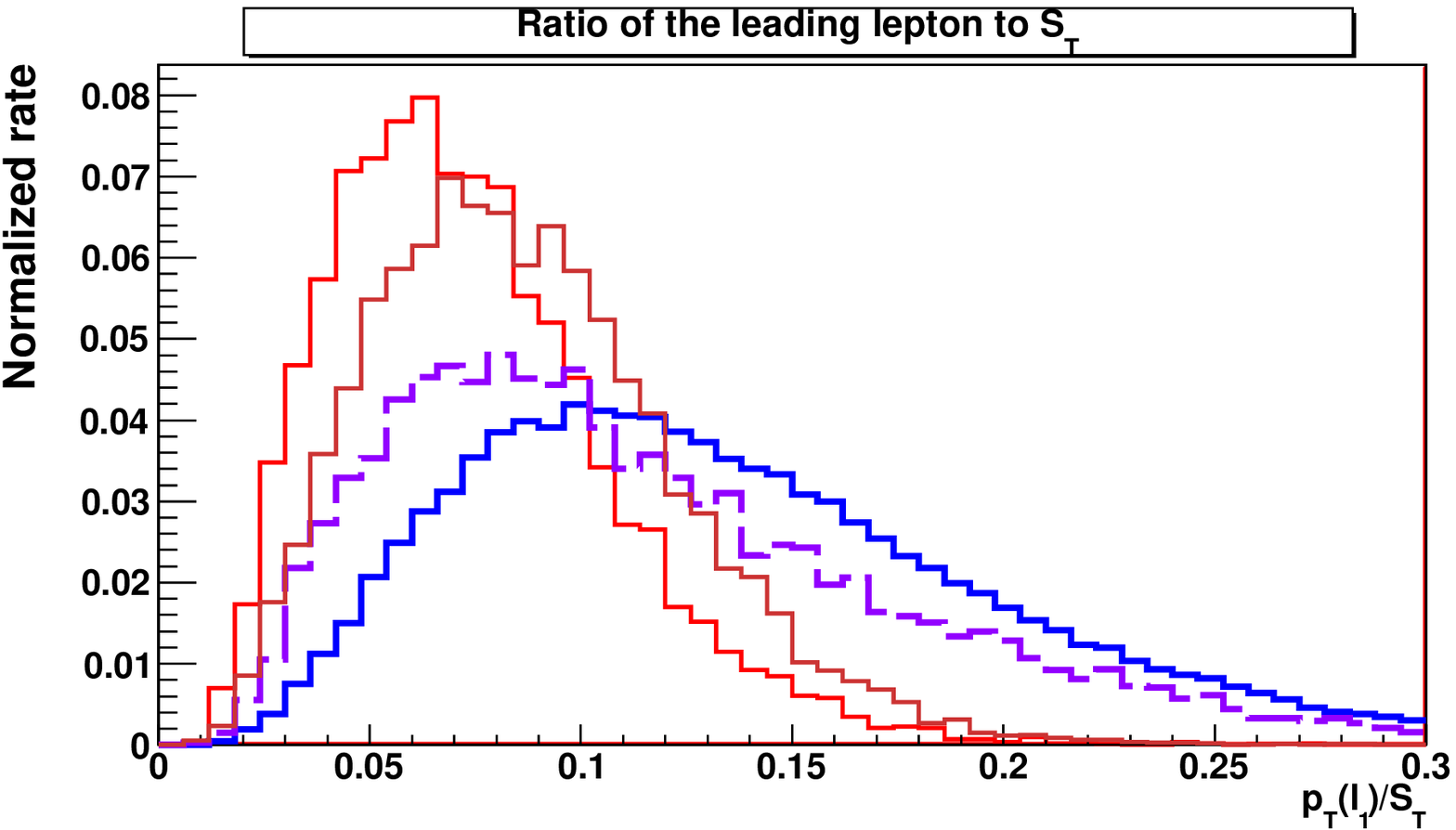}
\includegraphics[width=0.49\textwidth]{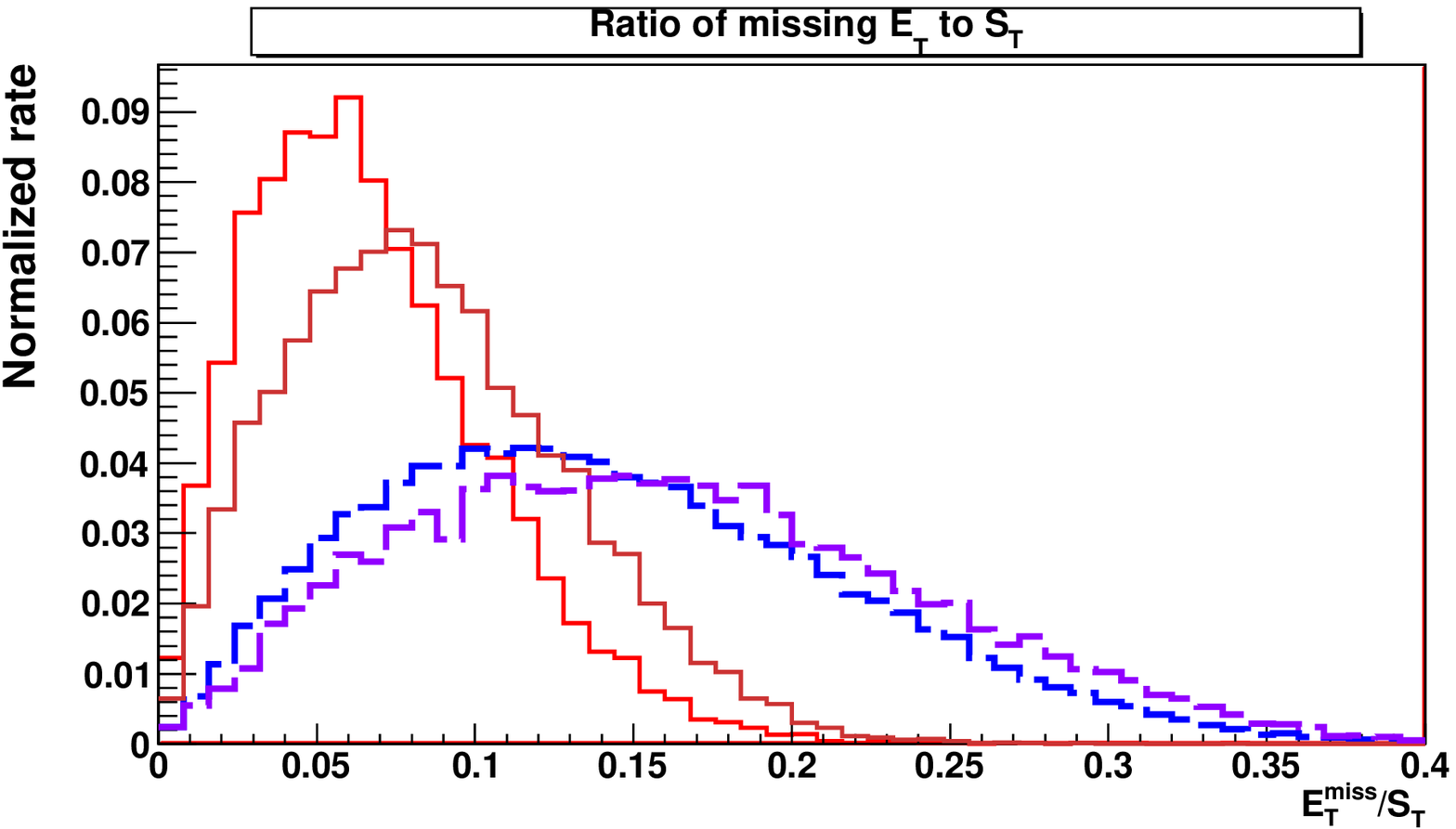}
\caption{Distribution of $r_l$ (on the left) and $r_\met$ (on the right) as defined in Eq.~\eqref{eq:newvar}. Light red and dark red curves stand for the signal spectra 1 and 2 (see Table~\ref{tab:benchWdecay}), blue curve represents the dileptonic background and violet represents $l \tau_l$ background. See Sec.~\ref{sec:reach} for details of simulations.}
\label{fig:ptl1_met_relative}
\end{figure}

It turns out one can do even better than just cutting on a high $\met$ and high $p_T(l_1)$ events. As we explained, the key feature of the $t \bar t$ events is that usually the leptons and the $\met$ are correlated with the hardness of the event, or $S_T$ defined as
\beq\label{eq:stdef}
S_T \equiv \sum_i p_T(j_i) + \sum_k p_T(l_k) + \met~.
\eeq  
On the other hand in the signal events these quantities are mostly uncorrelated. For this purpose we define the following variables:
\beq\label{eq:newvar}
r_l \equiv \frac{p_T(l_1)}{S_T}, \ \ \ r_{\met} \equiv \frac{\met}{S_T}~.
\eeq 
One should also prefer using these variables rather than $\met, \ p_T(l_1)$ because they are dimensionless and therefore cutting on them we do not introduce an explicit scale to the problem. 
We expect these quantities in the signal events to be in general small. We plot these variables for signal and background events in Fig.~\ref{fig:ptl1_met_relative} and it follows this expectation. Moreover, we see that $r_l $ and $r_{\met} $ are slightly less dependent on the particular spectrum than $p_T(l_1) $ and $\met$. In the next section we show that using this strategy together with the cuts on variables~\eqref{eq:newvar} we will have an excellent reach  after the $\sqrt{s} = 8$ TeV, $\cL = 20$~fb$^{-1}$ run.

 To summarize, the dileptonic channel is an excellent channel for the charged current decays. We will elaborate on a feasibility of this search explicitly and make more comments on the background behavior and shapes in section~\ref{sec:reach}.  

Finally we briefly comment on semileptonic and all-hadronic decay modes. The latter will probably be very hard to utilize, since it just results in multijets (up to 8 or even more) events without any evident handles like isolated leptons or missing $E_T$. In the semileptonic search one has signal events with isolated lepton, moderate $\met$ and at least 6 jets, typically resulting in small $\met $, high $H_T$ events. We will not try to elaborate on the feasibility of the cut-and-count search in this channel,  because evidently these searches are not optimal (they cannot reconstruct the resonances in jetty channels, taking advantage of the most interesting feature of the RPV signal) and basically already exist in some form both in ATLAS and CMS collaborations (they do not yield any interesting bounds though). It would be interesting to see though how the variables~\eqref{eq:newvar} can be used in these searches to improve further the reach and suppress the backgrounds.         

\begin{table}[t]
 \centering
\begin{tabular}{|c|c|c|c|}
\hline
  & $m_{\tilde b}$ & $ m_{\tilde t_1}$ & $\sigma (\tilde b \tilde b^*)$ \\
\hline 
\hline 
1 & 250 GeV & 186 GeV & 5.7 pb\\ \hline
2 & 270 GeV & 189 GeV & 3.7 pb \\ \hline
3 & 300 GeV & 217 GeV & 2.0 pb \\ \hline 

\end{tabular}
\caption{Benchmark points for the charged-current decay search. The production cross sections are given only for sbottoms .}
\label{tab:benchWdecay}
\end{table}

\section{Details of event simulations, backgrounds and reach}
\label{sec:reach}

To estimate the feasibility in dileptonic channel as explained in section~\ref{sec:signals} we analyzed three benchmark points points with masses presented in Table~\ref{tab:benchWdecay}. For the signal we simulated parton-level events with {\tt MadGraph 5}~\cite{Alwall:2011uj} for three signal benchmark points given in Table~\ref{tab:benchWdecay} and showered and hadronized them with {\tt Pythia~8}~\cite{Sjostrand:2007gs}. We wrote down a tailored model in {\tt Feynrules}~\cite{Christensen:2008py} for {\tt MadGraph~5} to capture the effects of the simplified model described in Sec.~\ref{sec:theory}. For the background we simulated the events in {\tt MadGraph 5} and showered with {\tt Pythia 6}~\cite{pythiamanual}. In order to capture correctly the effects of extra-jets (which are crucial for our analysis) we matched our samples up to two jets with the MLM procedure at 55 GeV. Events were clustered with {\tt FastJet~3}~\cite{Cacciari:2005hq,Cacciari:2011ma}.\footnote{Detector effects are neglected, but the results are sharp enough to survive full treatment.} 

\begin{figure}[t]
\centering
\includegraphics[width=0.49\textwidth]{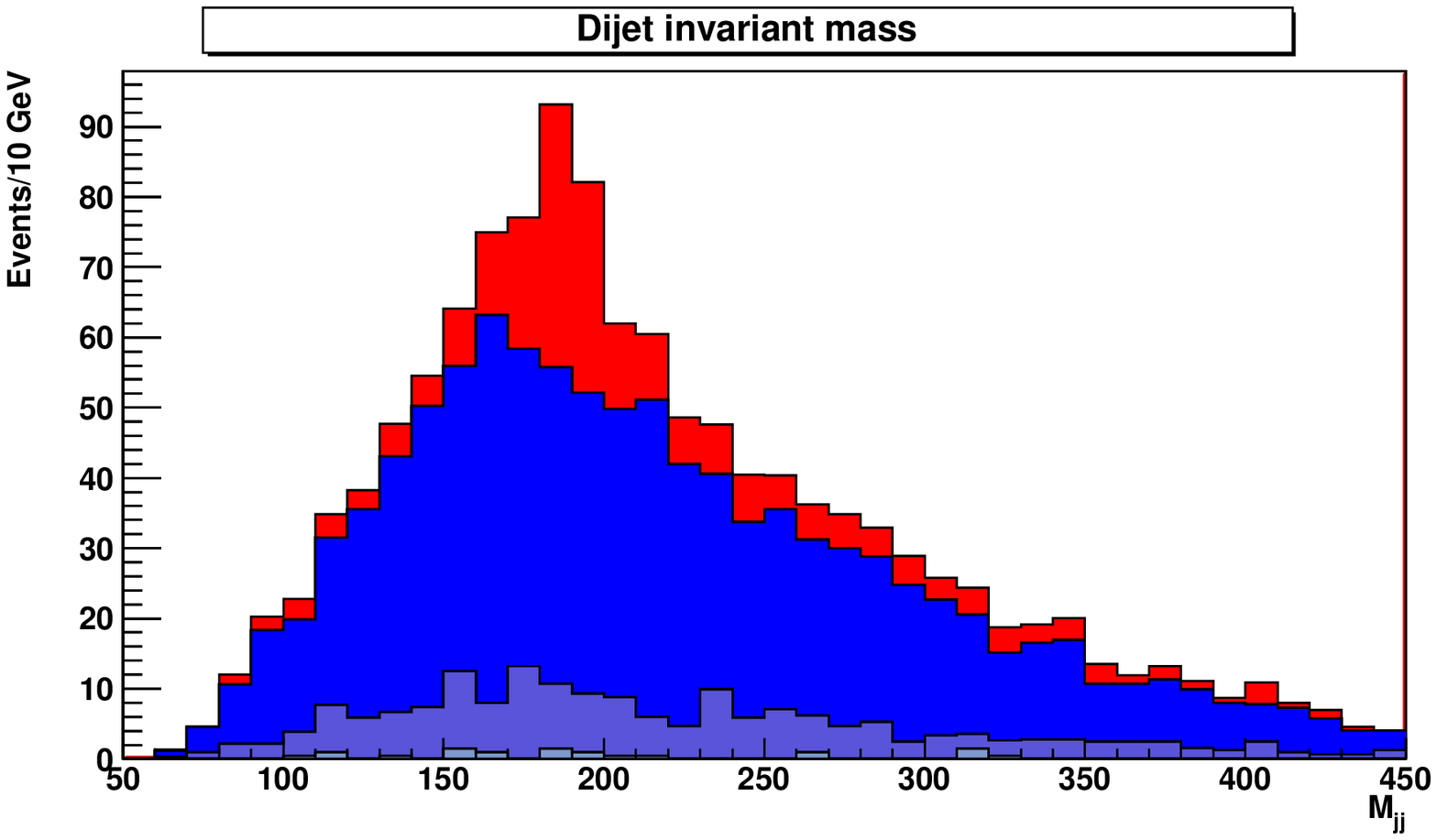}
\includegraphics[width=0.49\textwidth]{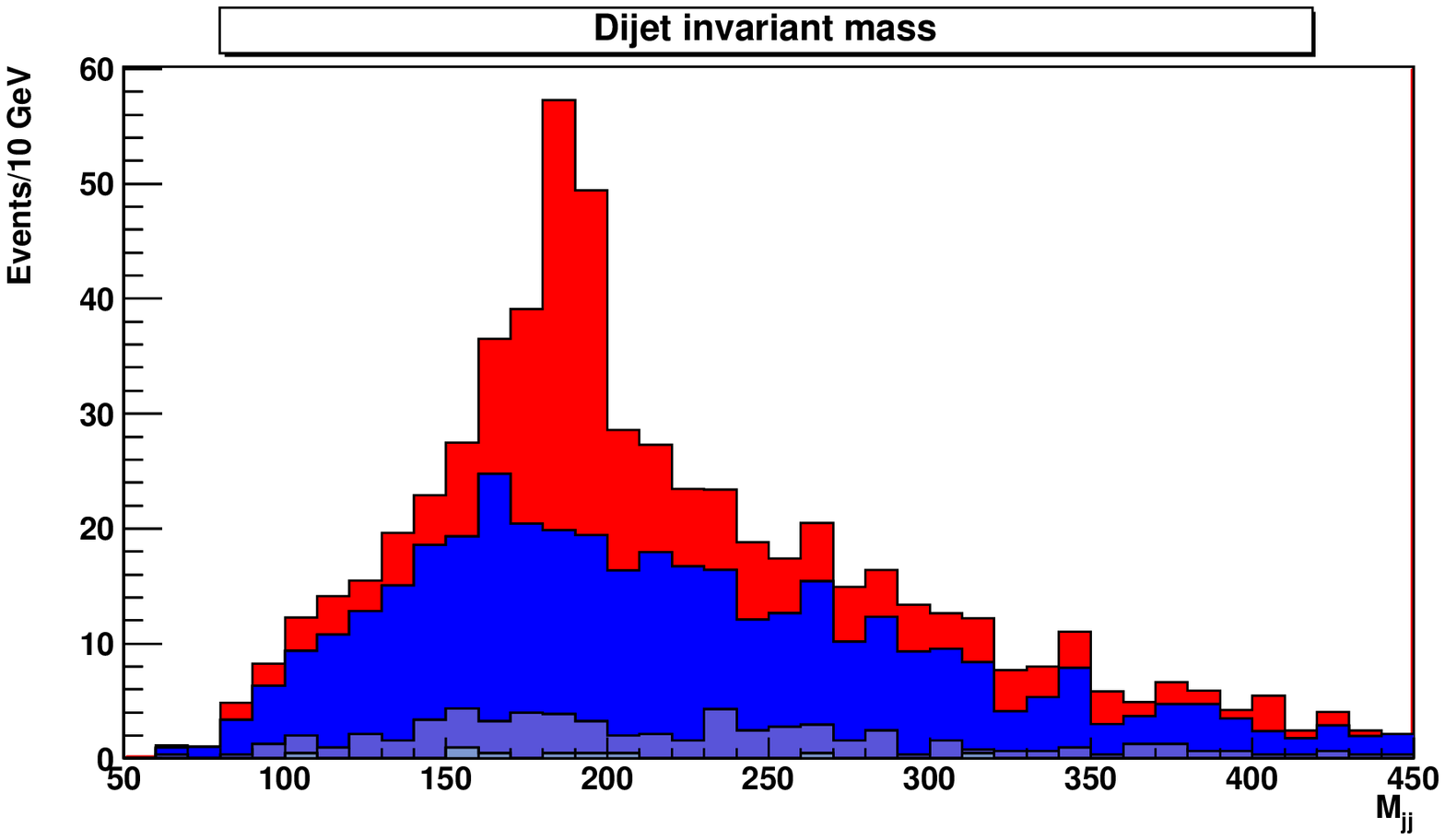}
\caption{Signal and background events for the benchmark point 1 after $\cL = 20 $ fb$^{-1}$. Red represents the signal, blue the dileptonic $t \bar t$ background, violet is $t \bar t, \ l\tau_l$ background and grey is  $t \bar t, \ \tau_l\tau_l$ background. On the LH side plot we do not impose b-veto, while on the RH side plot we do. We conservatively assume b-tag efficiency $\sim 40\%$. }
\label{fig:spec1}
\end{figure}

Following the discussion in section~\ref{sec:signals} we reconstruct our events and impose the following cuts:
\begin{enumerate}
 \item Cluster all the hadronic activity with anti-$k_T$ algorithm, clustering radius $R = 0.7$. Relatively large clustering radius is dictated by the fact that we are looking for the resonances, and smaller radius usually leads to losing relevant hadronic activity. The clustering radius is not optimized, but radii of order $R \sim 1.0$ are likely to be the most adequate.    
 \item Demand precisely two isolated leptons (carrying more than 85\% of the $p_T$ in the cone around the lepton with radius $R = 0.3$) in each event. We demand $p_T(l_1) > 20 $ GeV and $p_T(l_2) > 10$ GeV.\footnote{The logic of the cut on the $p_T$ of these leptons is dictated by trigger demands. Unfortunately the trigger information is not public. However relying on the logic of $\sqrt{s} = 7$ TeV run, we hope that the events with these leptons should be triggered on with sufficiently high efficiency, namely more than 90\%~\cite{Chatrchyan:2012ye}. Parenthetically we notice that if the threshold on the $p_T$ of the leading lepton can be lowered, the results that we performed can be further improved. Moreover, some of the events can be triggered on because they have sufficient $H_T$ or 4 or more sufficiently high-$p_T$ jets. We do not try to take into account the events which do not pass these lepton requirement, however lots of them can be ``salvaged'' since they pass other triggers and the ideal search will have to combine several different triggers.} The leptons should have $|\eta| < 2.5$. We discard the event if the leptons have same flavor \emph{and} $81 \ {\rm GeV } < m_{ll} < 101 \ {\rm GeV}$ to remove the background from Z + jets events.
\item Demand that the event is  sufficiently hard, $S_T > 400$ GeV as defined in Eq.~\eqref{eq:stdef} and $\met > 35$ GeV.
\item Require four or more hard jets in the event with $p_T(j_4) > 30 $ GeV. This requirement is natural since we are trying to reconstruct two resonances of $\tilde t_1$, which both decay into two quarks.
\item Using the variables in Eq.~\eqref{eq:newvar},  demand $r_{\met } < 0.15$ and $r_{l} < 0.15$.  
\item Try all possible pairings between four leading jets, and pick up the combination which minimizes the difference between the reconstructed invariant masses. Discard the event if the minimal possible mass difference is bigger than 10 GeV.\footnote{These cuts are not optimized, but it is also not very different from 7.5\% of the resonance mass which was used in~\cite{Aad:2011yh} . We explicitly checked our results with respect to variation of this cut. The results are rather stable as long as this cut does not exceed $\sim 25- 30$ GeV. We leave further optimization of these cuts to the experimentalists as it is also going to be affected by jet energy resolution.} If the event has five or more jets with $p_T > 25 $ GeV, try all possible pairings of two and three jets. If we get better results when taking the fifth jet into account, use the best combination which minimizes the mass difference between the reconstructed objects. 
\item Look for resonances in the reconstructed dijet invariant mass. 
 
 \end{enumerate}

\begin{figure}[t]
\centering
\includegraphics[width=0.49\textwidth]{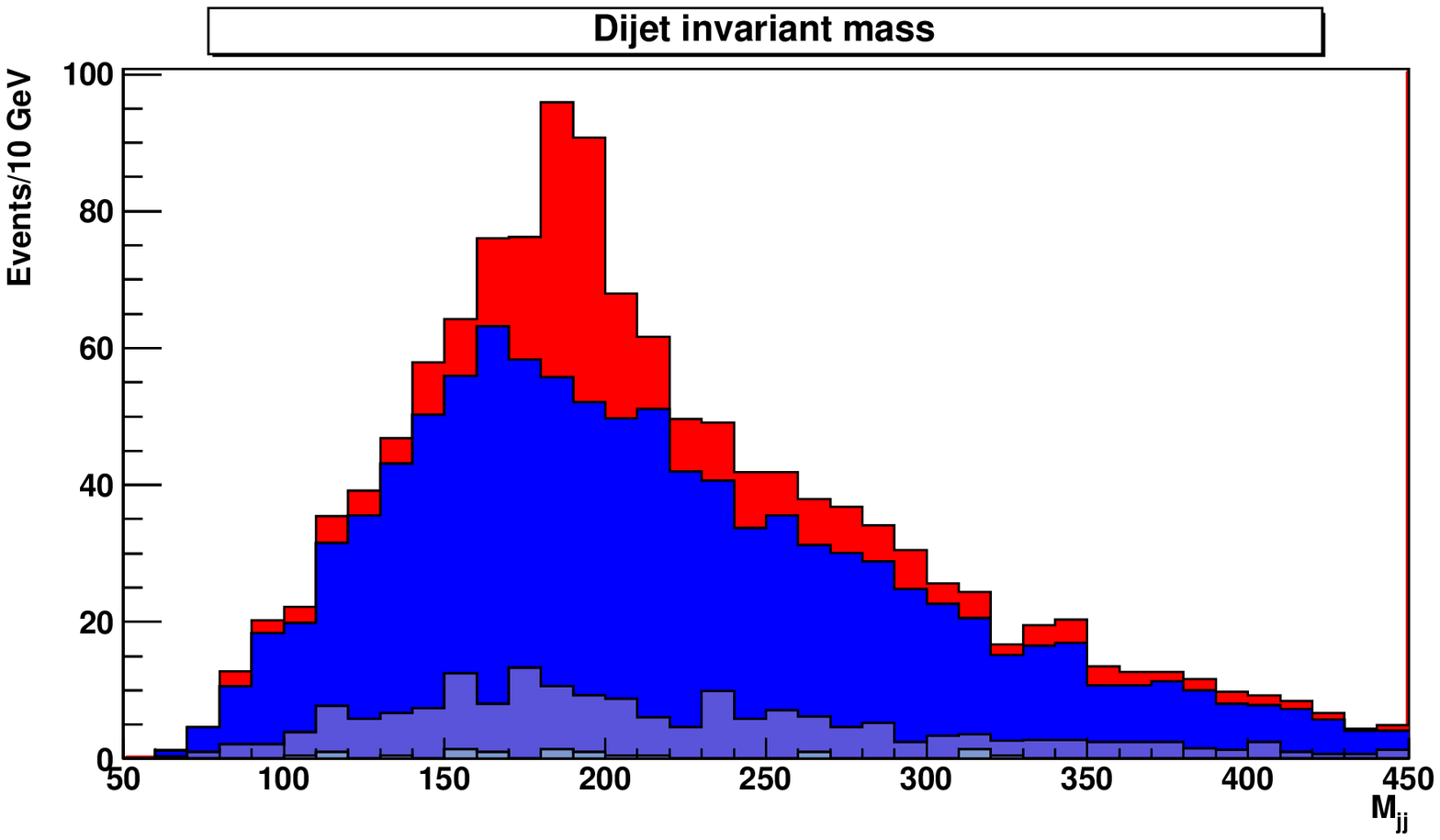}
\includegraphics[width=0.49\textwidth]{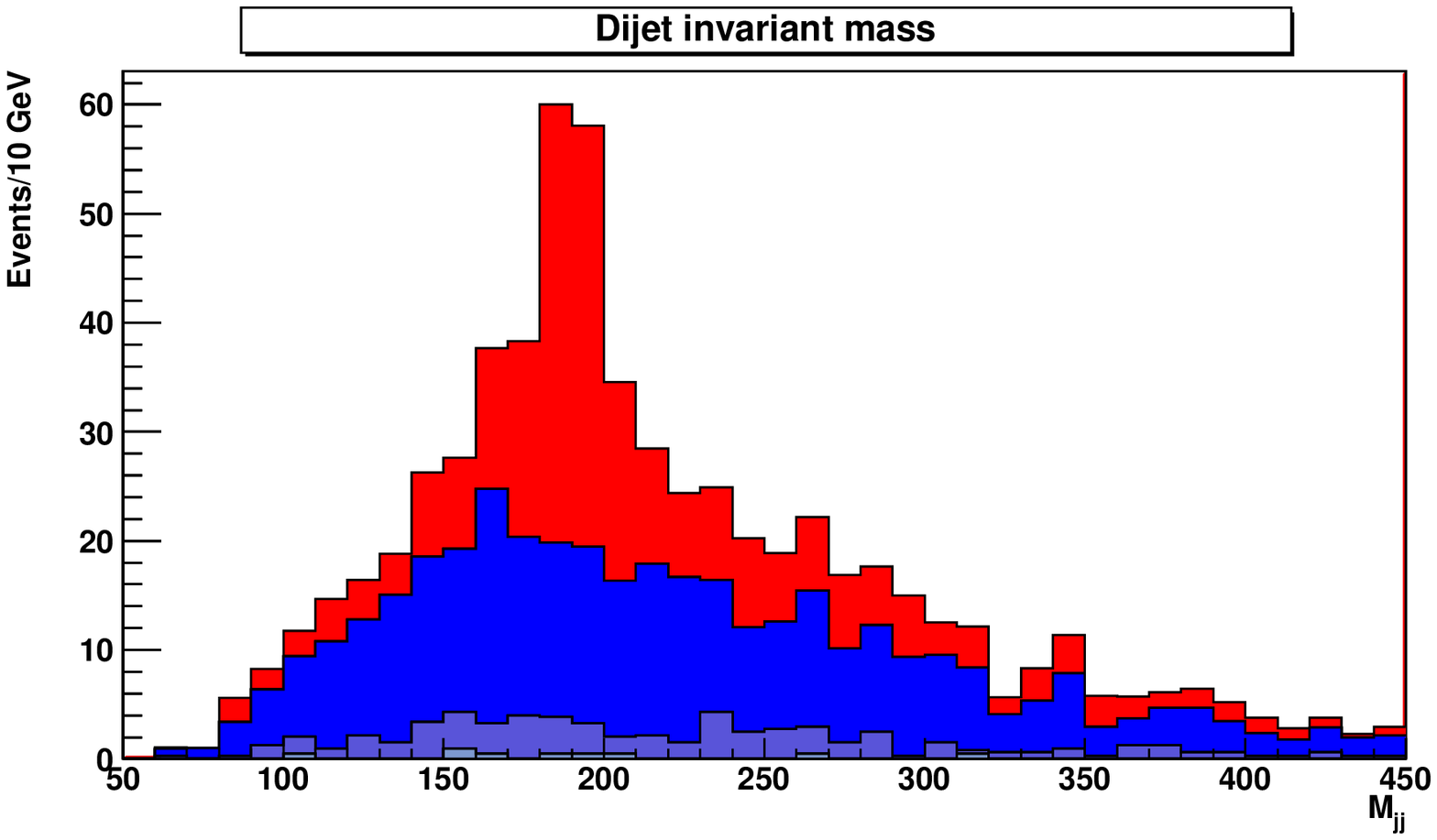}
\includegraphics[width=0.49\textwidth]{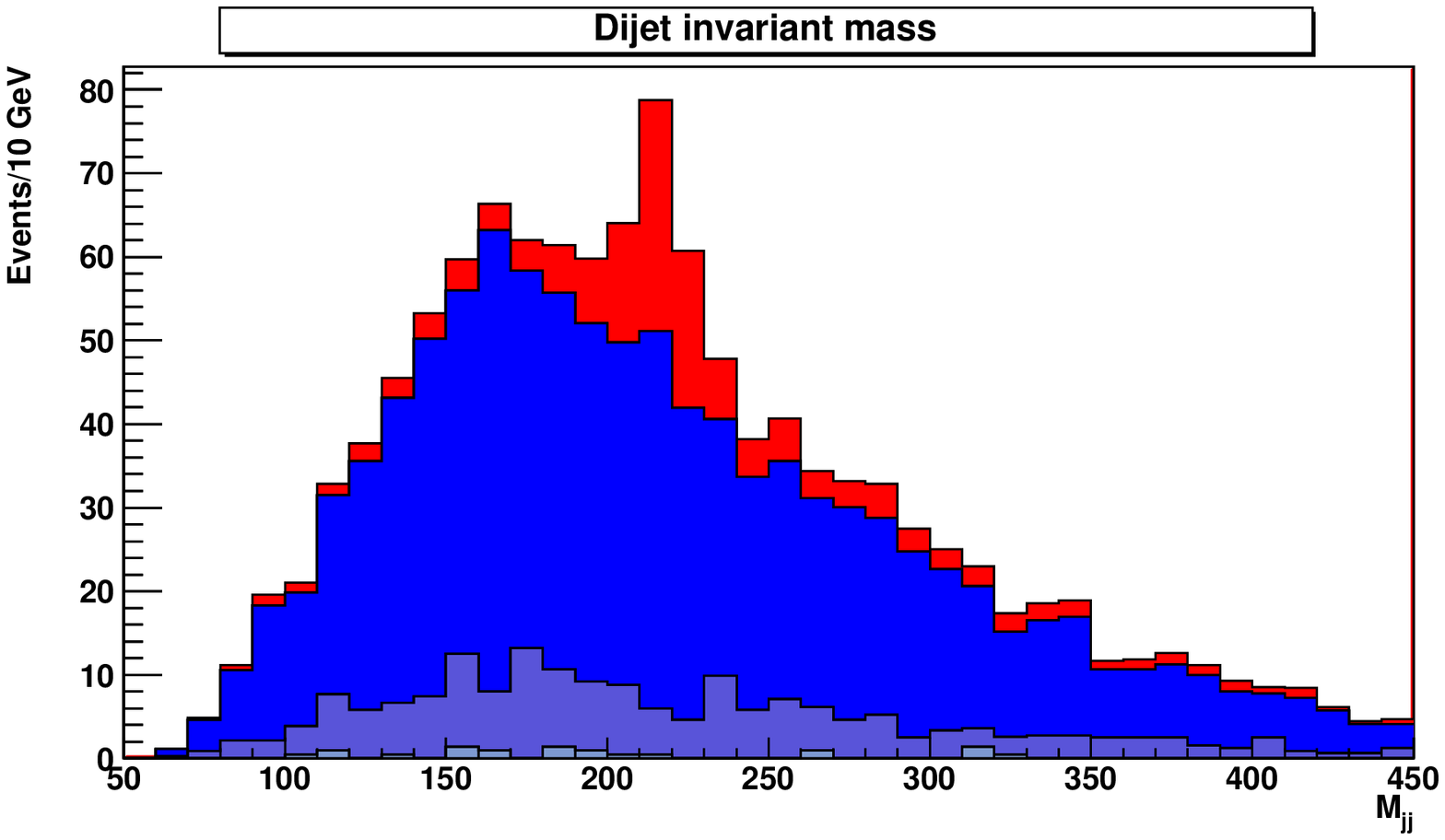}
\includegraphics[width=0.49\textwidth]{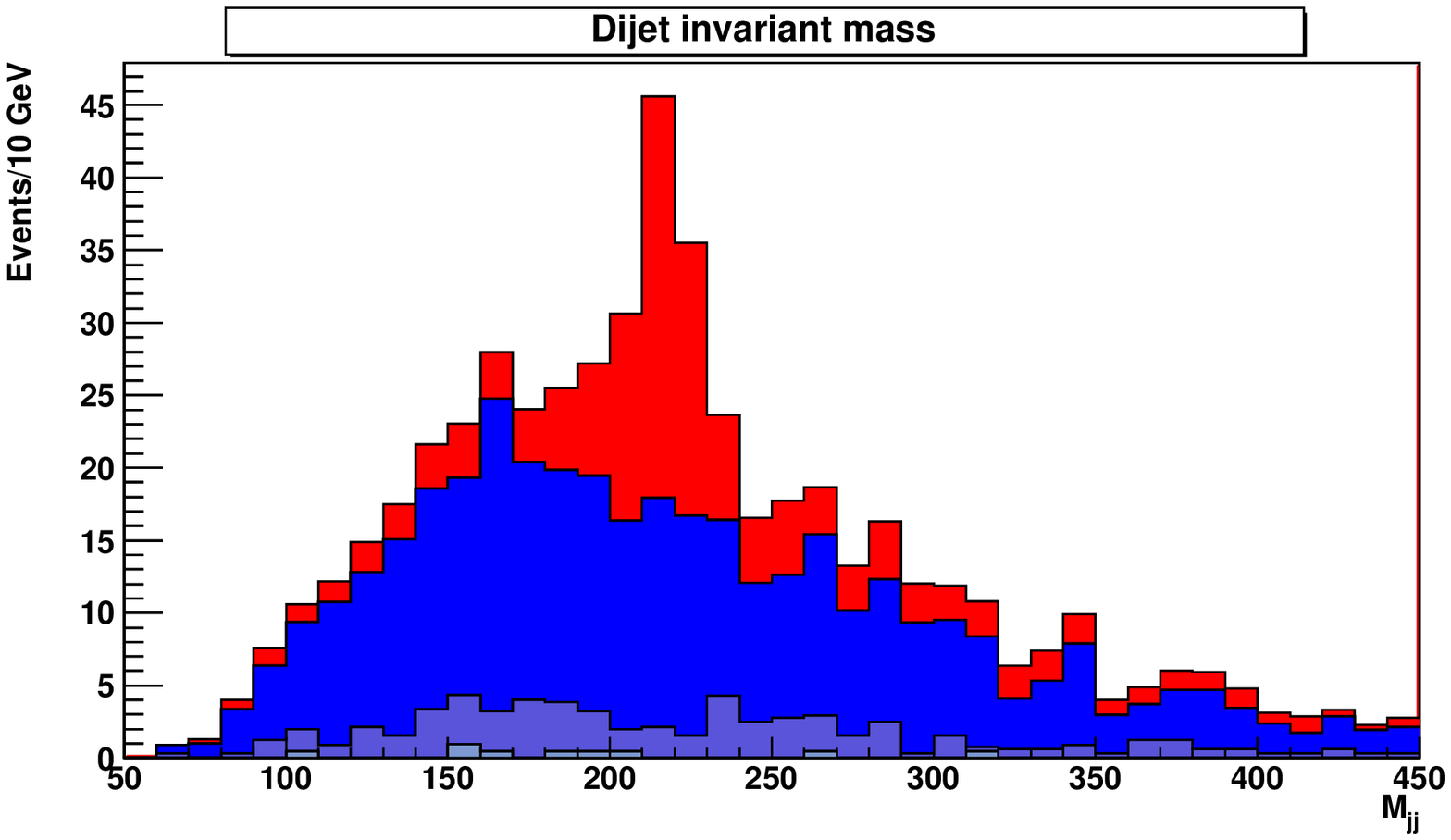}
\caption{Signal and background events for the benchmark points 2 (up) and 3 (down) after $\cL = 20 $~fb$^{-1}$. Red represents the signal, blue the dileptonic $t \bar t$ background, violet is $t \bar t, \ l\tau_l$ background and grey is  $t \bar t, \ \tau_l\tau_l$ background. On the LH side plot we do not impose b-veto, while on the RH side plot we do. We conservatively assume b-tag efficiency $\sim 40\%$. }
\label{fig:spec2_3}
\end{figure}

Before we present the results of our simulations we discuss the backgrounds to our analysis. Clearly the most formidable background is dileptonic $t \bar t$ (also including the leptons coming from leptonic $\tau$ decays). Naively, one could also worry about $Z\to (\tau_l \tau_l)~+$~jets, as well as DY $l^+ l^-$ production and $W^+ W^-+$~jets. We do not simulate these backgrounds and we rely on experimental results which found these backgrounds negligible to $t \bar t$ with the cuts which were very similar to ours. First, it was shown in~\cite{CMS:ttbarXsec} that the background $Z \to (\tau_l \tau_l)~+ $~jets becomes completely negligible to $t \bar t $ when the third hard jet is required in the event. We also see from Table~1 in~\cite{CMS:ttbarXsec} that the DY background is subdominant to $t \bar t$ at least by factor of 5 after requiring at least two hard jets and $\met > 35$ GeV. However we demand four hard jets in our events, which is supposed to decimate the DY dileptonic production and render it completely negligible to $t \bar t$. Therefore we will further concentrate on $t \bar t~+$~jets as the dominant background to our signal, and neglect the subdominant channels. 

Since we are looking for  bumps in the dijets invariant-mass distribution, it would first be helpful to understand what effects our cuts have on the backgrounds and how  they shape the background distribution. Not surprisingly, before all the cuts $m_{jj}$ in the background is a smoothly falling distribution which is peaked around 50 GeV (this peak is carved by our demand from each jet to have  $p_T > 25$ GeV. Further demands on hardness of the event move this peak to significantly higher values of masses. For example a cut on $H_T\equiv \sum_i p_T(j_i) > 400 $ GeV (which does a reasonable job with suppressing the backgrounds) moves this peak to the vicinity of 200 GeV, which is uncomfortably close to the mass scale where we are looking for our resonances. Cuts on the $S_T$ and the $p_T$ of the softest necessary jet have a similar effect, therefore we choose the cut in points~(3) and~(4) discussed above to be relatively moderate (one could choose way harder cut which would still remove more background than signal events, for the price of moving the peak of the background distribution to higher masses). On the other hand, cuts on $r_\met $ and $r_l$ in point~(5) do not have this effect, they relatively uniformly discriminate against background from all the invariant masses (as this distribution have already been shaped by $H_T,\ S_T$ and $p_T(j_4)$). Note also, that this terrain of models is still relatively unexplored and the new physics can hide at very low masses. It is very natural to expect that the lightest stop has a mass of $\cO (200\ {\rm GeV})$ or even lighter. Therefore we emphasize that given a choice of cuts, one should always prefer harsher cuts on $r_\met$ and $r_l$, preferring as mild as possible cuts on ``hardness variables'', namely $\met, H_T, S_T, p_T(j)$. This approach is ultimately dictated by our attempt to avoid carving spurious bumps on the background distributions. 

Armed with this understanding of the background behavior we turn to the actual analysis. For this purposes we assume the NLO $t \bar t $ production cross section at $\sqrt{s} = 8$~TeV to be 205 pb (this result is taken from MC@NLO~\cite{Frixione:2002ik,Frixione:2003ei} with default scale choice). Practically, the cross section for $t \bar t$  production is going to be slightly bigger, tallying up to 230~pb (from NNLL resummation) with $\cO(10\%) $ uncertainty~\cite{Beneke:2011mq,Cacciari:2011hy}. However we take NLO results to be consistent in our estimates comparing the signal, which we know at the NLO, to the background. As we will see our results are strong enough, that increasing the background moderately without changing the signal cross section by no mean changes our conclusions. We present the results of our simulations in Figs.~\ref{fig:spec1} and~\ref{fig:spec2_3}. In both cases we present the results with and without b-veto. As explained in Sec.~\ref{sec:theory} the lightest stop can have decay modes which either include a b-jet or not. Since the dominant background is $t \bar t$, one can achieve much better reach if the signal events contain only light quarks. In this case we perform b-veto which further reduces the background, leaving the signal intact. The plots on the RH side of Figs.~\ref{fig:spec1} and~\ref{fig:spec2_3} correspond to this picture. On the other hand, if the stop decays to a b-quark + a light flavor, we cannot perform the b-veto (LH side plots), but the backgrounds are still under very good control and one has a reasonable discovery reach for all three benchmark points. 

\begin{figure}[t]
\centering
\includegraphics[width=0.49\textwidth]{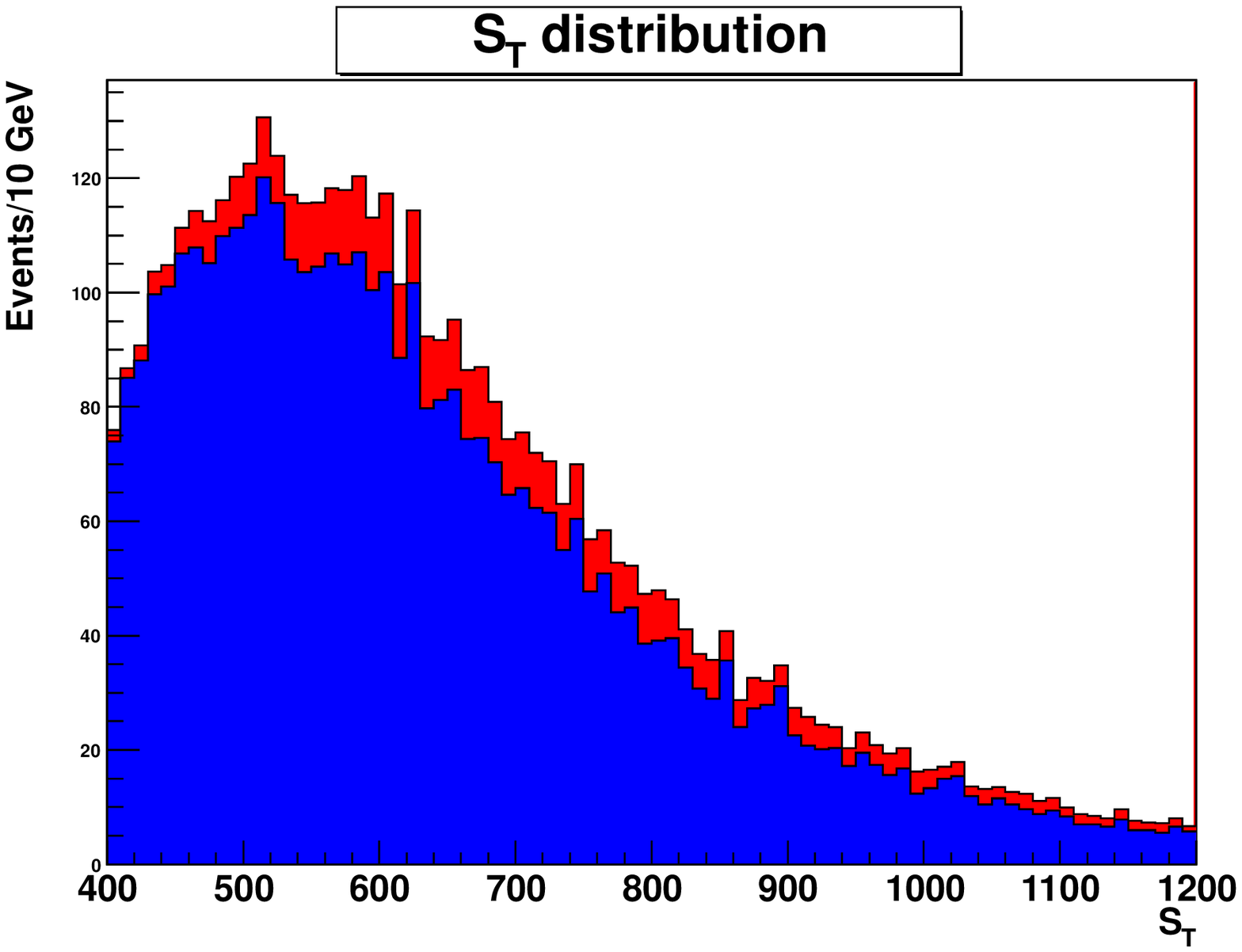}
\includegraphics[width=0.49\textwidth]{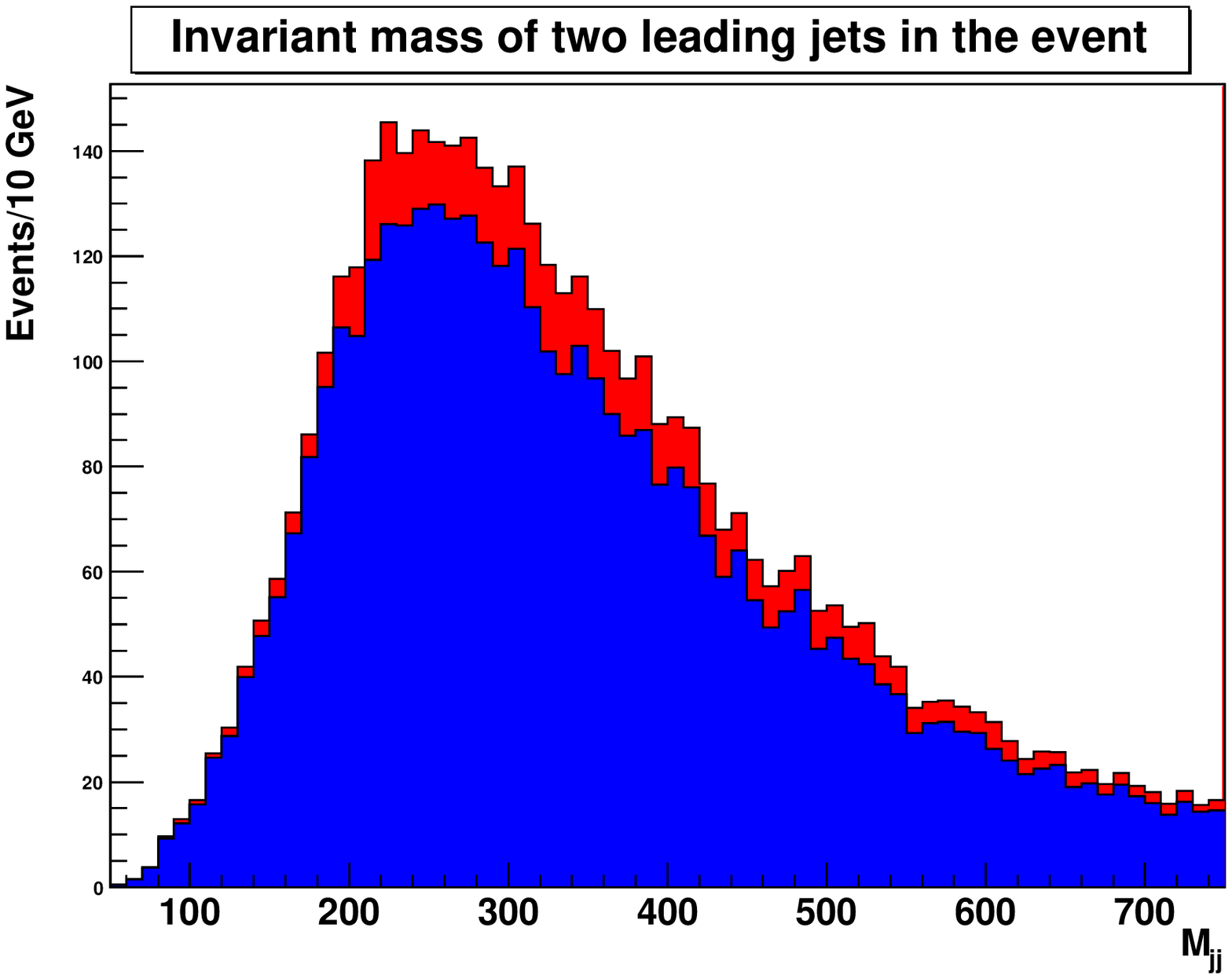}
\caption{Distribution of $S_T$ (left) and invariant mass of two leading jets invariant mass (right) in signal and background events. We use the benchmark point~(3) for these plots. Only dileptonic $t \bar t$ background is plotted. We use all the same cuts as in resonance search \emph{except the point~(6)}. Using these distributions as discriminator in a cut-and-count search will be at least very challenging.  }
\label{fig:wrong_strategy}
\end{figure}

Finally, we briefly explain why our resonance search in this channel is much more efficient than simple cut-and-count searches.  One could naively expect that the discriminators that we use to separate the signal from the background are sufficient for an easier cut-and-count search. This naive expectation is not true. To illustrate this we plot in Fig.~\ref{fig:wrong_strategy} distributions of $S_T$ and the invariant mass of the two leading jets. We use all the same cuts as in our resonance-searching analysis except the step~(6). We see in Fig.~\ref{fig:wrong_strategy} that using a cut-and-count strategy will be, in the best case, extremely challenging and will require detailed understanding of the \emph{normalization of the background}. Therefore we conclude that di-resonance search is the optimal strategy here.     

\section{Brief Comments on Neutral Current Decays}
\label{sec:neutral} 
\begin{figure}[t]
\centering
\includegraphics[width=0.9\textwidth]{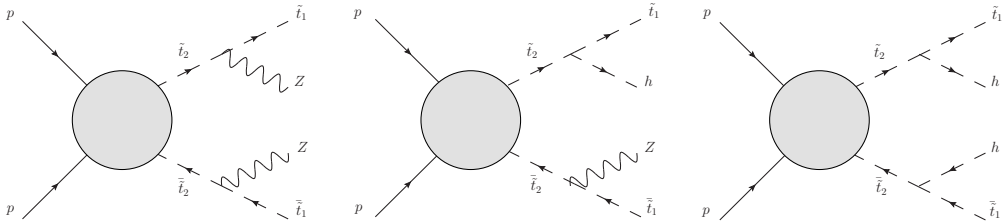}
\caption{Neutral current decay processes which can be relevant for the most minimal spectrum.  }
\label{fig:neutral_current}
\end{figure}

Until now we were very detailed in describing the searches for charged-current decays. However the charged current decays reveal only part of the full picture. As we explained in Sec.~\ref{sec:theory}, in certain spectra the charge-current decay will dominate the collider signatures, while in other spectra, the neutral current decays will leave the most spectacular signatures (see processes~(2) and~(3) in Eq.~\ref{eq:processes}). We will briefly comment on these processes (see Fig.~\ref{fig:neutral_current} for a summary of diagrams)  in the current section, but we will be less detailed because as we will see the most important potential discovery channel (the multilepton channel) is already considered by the CMS. Other channels are not expected to be as strong as mutileptons and will mostly favor cut-and-count strategy rather than resonance reconstruction due to the high multiplicity of jets.

In  neutral-current decay events we have pair-produced heavy stops decaying into their light partners emitting two $Z$'s, two Higgses, or one $Z$ and one Higgs. $Z$'s can be emitted  either on- or off-shell, while Higgs decays are very unlikely to proceed off-shell due to very strong bottom Yukawa suppression. Higgs decays are rarely spectacular, the most important Higgs decay mode (assuming that its mass is $\sim 125$ GeV, as current experimental hints suggest) is $h \to b \bar b$ and the third important is $h \to gg$. If the heavy stop decays to the light one emitting two additional jets (b-quarks or gluons) we get a very challenging event without any obvious handles, i.e. without MET and/or leptons. The second important Higgs decay mode $h \to WW^*$ (BR bigger than 20\%) and the fourth important ($h \to \tau \tau$) are more distinctive, but the resulting  going rate of the higgs into two leptons (either through $W$ or through $\tau$) is smaller than the rate of $Z$. Therefore we will mostly concentrate on neutral current decays with the $Z$ in final state.   

Both $Z$'s decaying leptonically can probably be considered a ``golden'' channel, even though  the branching fractions are very small, $\sim 0.5\%$. These rare but very spectacular events with up to four isolated leptons and lots of energetic jet activity can be probed by multilepton searches. The backgrounds for these events from the SM are extremely small and therefore even observation of few events can be considered discovery (or, alternatively, even the benchmark points which are expected to yield very few events in certain channels can be excluded). 

To illustrate this point we explicitly compared a yield of three different benchmark points with the results of analysis~\cite{Chatrchyan:2012ye}. This CMS analysis is very special because it has a very low $p_T$ threshold for the leptons (leptons, which are as soft as $p_T = 8$ GeV  are considered, higher $p_T$s are required for trigger leptons though). The expected yields of all three different points are presented in Table~\ref{tab:multleps}. We also compare the yields with the theoretical expectations and experimental results, which are quoted from~\cite{Chatrchyan:2012ye}. In all these points we assumed, for simplicity of estimation, 100\% branching ratio for $\stp_2 \to \stp_1 Z$ decay. However, more realistically, such decays would account for only an order one (but highly spectrum-dependent) fraction of $\tilde{t}_2$ decays. 
If higgsinos lie between the two stops, than $\stp_2 $ decay will likely be dominated by $\stp_2 \to b \tilde H^+$ (mediated by top Yukawa coupling). Moreover, as discussed in section~\ref{sec:theory}, the sbottom should lie between the stops, so that $\tilde{t}_2$ should also undergo charged current decays to sbottom. For squeezed enough spectrum such decays may produce even softer leptons than the neutral current decays, and could be more challenging to detect. For the estimates below, we ignore such subtleties and study 
the ansatz of neutral-current dominance. We will see that with enough statistics, even with significant depletion to other channels, neutral current decays might provide spectacular evidence for supersymmetry. 

Clearly with this assumption of 100\% branching fraction to $Z$ the first two benchmark points in Table~\ref{tab:multleps} are excluded. However full exclusion plots are beyond the scope of this paper, because they would demand more refined simulations and more well defined assumptions about the Higgsinos, sbottoms and the mixing angle between the stops. Note that if the mass splitting between the stops is smaller than 125 GeV the decay will almost never proceed through $h^*$  due to smallness of bottom Yukawa. Alternatively, pushing the heavy stop mass above 300 GeV, we begin to lose the sensitivity in this channel, therefore we find that there is a big part of parameter space which is far from exclusion and probably is not expected to make a significant contribution to the multilepton channel even after $\cL = 20$~fb$^{-1}$ run. 

The yields are still smaller than one would naively expect. Most of the events either do not have an isolated leading lepton harder than 20 GeV, or the next to leading lepton harder than 10 GeV, which renders them unsuitable for a dileptonic trigger. As expected the most populated bins are those with low $\met$ and high $H_T$, the $H_T$ in our events comes from $\stp_1 \to jj$ decays, while the MET is merely instrumental. The prediction in $4l$, high $H_T$ low MET channel are already in tension with the experiment because of extremely low (essentially non-existent) SM background.   

It is also expected that there is relatively high chance to lose at least one of the leptons due to isolation criteria, or due to high rapidity, so the bins with three leptons turn out to be even  more informative than the bins with four isolated leptons. Interestingly, the only bin where we could predict a significant excess, low MET and high $H_T$ without $Z$, is precisely the bin where CMS observes a non-negligible  excess of events, recording 11 events where $4.5 \pm 1.5$ events are expected. Again, the yield of our benchmark points is too high to explain the excess, unless non-vanishing (but also not overwhelming) BR for $\stp_2 \to \tilde H^+ b$ is assumed. We do not try to claim that it is an anomaly, or that we try to explain this excess, however it is probably an interesting channel to watch when $\sqrt{s} = 8$~TeV data is analyzed. We do not try to estimate the reach of the multilepton channels for $\sqrt{s} = 8$~TeV  since the event yield both of the signal and the background is very low, and the backgrounds are very hard to estimate. Nonetheless it is clear that the LHC right now is on the edge of probing an interesting region, and high $H_T$ low MET channels are of particular interest to watch.         

\begin{table}[t]
\centering  
\begin{tabular}[t]{|c||c|c|c|| c|c|c|}
\hline 
 Spectrum & \scriptsize $m_{\stp_1} = 180 $ GeV & \scriptsize $m_{\stp_1} = 185 $ GeV &  \scriptsize $m_{\stp_1} = 189 $ GeV & Exp. & Err.& Obs. \\
Selection& \scriptsize $m_{\stp_2} = 245 $ GeV & \scriptsize $m_{\stp_2} = 260 $ GeV &  \scriptsize $m_{\stp_2} = 277 $ GeV & \cite{Chatrchyan:2012ye} & \cite{Chatrchyan:2012ye} & \cite{Chatrchyan:2012ye} \\
\hline 
\hline 
\scriptsize 4l $\met > 50,\ H_T > 200$ & $< 0.1$ & $< 0.1$ & $< 0.1$ & 0.018 & 0.005 & 0  \\
\hline 
\scriptsize 4l $\met > 50,\ H_T < 200$ &  $<0.1$ & $< 0.1$ & $< 0.1$ & 0.2 & 0.07 & 1  \\
\hline 
\scriptsize 4l $\met < 50,\ H_T > 200$ & 3.6 & 3.5 & 1.8 & 0.006 & 0.001 & 0 \\
\hline 
\scriptsize 4l $\met < 50,\ H_T < 200$ & 1.0 & 0.6 & 0.4 & 2.6 & 1.1 & 1  \\
\hline 
\scriptsize 4l $\met > 50,\ H_T > 200$, Z & $< 0.1$ & $< 0.1$ & $< 0.1$ & 0.22 & 0.05 & 0 \\
\hline 
\scriptsize 4l $\met > 50,\ H_T < 200$, Z & $< 0.1$ & $< 0.1$ & $< 0.1$ & 0.79 & 0.21 & 1 \\
\hline 
\scriptsize 4l $\met < 50,\ H_T > 200$, Z & 1.0 & 1.0  & 2.5  & 0.83 & 0.33 & 1\\
\hline 
\scriptsize 4l $\met < 50,\ H_T < 200$, Z & 0.3 & 0.2 & 0.4 & 37 & 15 & 33 \\
\hline 
\scriptsize 3l $\met > 50,\ H_T > 200$ & 0.3 & 0.4 & 0.7 & 5.0 & 1.3 & 8 \\
\hline 
\scriptsize 3l $\met > 50,\ H_T < 200$ & $< 0.1$  & $< 0.1$ & $< 0.1$ & 27.0 & 7.6 & 30 \\
\hline 
\scriptsize 3l $\met < 50,\ H_T > 200$ & 16.9 & 14.1 &  8.8 & 4.5 & 1.5 & 11 \\
\hline 
\scriptsize 3l $\met < 50,\ H_T < 200$ & $< 0.1$ & $< 0.1$ & $< 0.1$ & 144 & 36 & 123 \\
\hline 
\scriptsize 3l $\met > 50,\ H_T > 200$, Z & $< 0.1$ & $< 0.1$ & 0.2 & 18.9 & 6.4 & 20 \\
\hline 
\scriptsize 3l $\met > 50,\ H_T < 200$, Z & $< 0.1$ & $< 0.1$ & $< 0.1$ & 134 & 50 & 141 \\
\hline 
\scriptsize 3l $\met < 50,\ H_T > 200$, Z & 2.5 & 2.3 & 4.5 &19.2 & 4.8 & 15 \\
\hline 
\scriptsize 3l $\met < 50,\ H_T < 200$, Z & $< 0.1$ & $< 0.1$ & $< 0.1$ & 764 & 183 & 657 \\
\hline 
\end{tabular}
\caption{Expected yields of events in the ``golden'' channel in the multilepton search of~\cite{Chatrchyan:2012ye}, $\sqrt{s} = 7 $ TeV with assumption ${\rm BR} (\stp_2 \to Z^* \stp_1) = 100\%$. Channels with high $H_T$, low MET are the most informative. All possible leptonic decays of $Z^*$ have been simulated including leptonic $\tau$s. We define the $Z$ window such that the invariant mass of the OSSF pair is $76\ {\rm GeV} < m_{ll} < 106$ GeV. We do not simulate channels with hadronic $\tau$s due to difficulties to mimic one-prong $\tau_h$ detection with our theoretical tools. Three right columns cite the results of~\cite{Chatrchyan:2012ye} where Exp. stands for the expected yield of the SM, Err. is the systematic error as it was estimated by the experimentalists, and Obs. stands for the observed number of events at $\cL = 4.98$ fb$^{-1}$.  }
\label{tab:multleps}
\end{table}      

The second important channel in this category, is one where one of the $Z$'s decays leptonically, while its counterpart decays invisibly. This channel has higher branching ratio of order 3\% but has bigger backgrounds. It also has signatures which naively resemble R-parity conserving SUSY - namely opposite-sign dileptons with jets and $\met$. If the mass gap between the stops is sufficiently big, such that $Z$ decays on-shell, the signature is leptonic $Z$ + jets + $\met$, naively resembling one of the well-known signatures of R-parity conserving gauge-mediation~\cite{Meade:2009qv}. Alternatively, if the $Z$ decays off-shell we find opposite-sign same-flavor pairs and $\slashed E_T$, again very similar to a standard R-parity conserving signature with decay chain proceeding through a low mass slepton.

Unfortunately these resemblances are not close enough to be useful.  For example, we explicitly checked the event yield for all three reference points in table~\ref{tab:multleps} in analysis~\cite{Aad:2011cwa} and found that the yields are  far below  values which one needs in order to have exclusion (usually yielding one event or even less in each of the signal regions). These channels are so different because they typically have very low missing $E_T$ and relatively soft leptons, which makes the discovery very difficult with simple cut-and-count experiments. 

Nonetheless one can exploit these events if different strategies are used. If the $Z$ goes off-shell we get precisely a signature which is identical to the charge-current decay in the dileptonic channel (see Fig.~\ref{fig:feyndiag}) and was analyzed in details in Secs.~\ref{sec:signals} and~\ref{sec:reach}. One can use precisely the same techniques and if an excess in multilepton is found and confirmed in the dilepton + MET channel, this  can be an excellent cross-check to establish if the excess indeed comes from $\stp_2 \to Z^* \stp_1$ decay chains.       

One can also suggest similar searches when $Z$ decays on-shell. In this case the search should be modified, because removing the events in the $Z$-window would wash our signal out. Maybe this search is also feasible, however with theory tools it will be hard to estimate reliably the background which comes from $(Z\to l^+ l^-)+$ jets with instrumental $\met$. Therefore, we point out that this search can be tried, but we do not make any conclusions about the backgrounds. 

Although all other channels of $Z$-decays have much bigger branching ratios, we do not see any clear strategies for how these can be utilized. The case where one $Z$ decays leptonically and the second hadronically would suffer from an enormous $l^+ l^-$ DY background (if $Z $ is on-shell, we get  $Z+$~jets, which is even worse), without even modest MET. One faces a similar problem if one of the $Z$'s decays invisibly and the second one decays hadronically.

\section{Conclusions and Outlook}
\label{sec:conclusions}

The main results of this paper are new searches that we propose for the 8 TeV LHC and the novel techniques that we find useful to discriminate the new physics signal from the background. These searches are motivated by natural SUSY with renormalizable baryon-number violating RPV interactions. This yields a set experimental signatures
which are not efficiently captured by current LHC searches. 
We pointed out that the signatures of charged-current decays can be discovered by a  di-resonance search with two additional leptons. The leptons, being soft and accompanied by modest MET are essentially useless for cut-and-count search, but provide us an excellent handle and allow us to see the dijet resonances despite small production cross section. It is in fact surprising how efficient these searches can be. Moreover, there is a good reason to believe that one can do even better than our estimates. Although we tried to choose an adequate  clustering radius, we neither optimized it nor used ``grooming'' techniques. Simple optimization and using  ``trimming''~\cite{Krohn:2009th}, which is the most adequate grooming technique for these purposes, can further improve the sensitivity. 

To efficiently discriminate the signal from the background (which is almost completely composed of $t \bar t$ events) we propose to use a set of rather novel cuts combined with more standard tools. On one hand we are cutting on the hardness of the MET, leptons and the entire event, which is a standard tool, but we also emphasize that these cuts should not be too harsh. On the other hand we propose to put an upper cut on $r_l$ and $r_{\met}$ variables, which is a novel way to discriminate signal events where the hardness of the special objects in the event (leptons and $\met$) is uncorrelated with the overall hardness of the event. 

The discriminators  that we propose to use are not completely unknown, for example CMS was using a much weaker version of $r_l$ as one of the variables in artificial neural network in their dileptonic analysis~\cite{CMS:ANN} (whose reach is hard to estimate because it uses a cumbersome multi-variate approach). The variable $r_\met$ has not yet been use in any analysis that we are aware of. We point out that use of these tools can go much beyond the particular analyses that we propose, and can be used in cut-and-count experiments as well as in resonance searches. We point out that these techniques are suitable in any new physics scenario where one finds transitions between the states with small mass splitting (see e.g.~\cite{Katz:2009qx,Katz:2010xg}).

Finally we point out that  the searches that we propose form one more important step in the program to map the collider signatures of RPV natural SUSY (for previous works see~\cite{Kilic:2011sr,Allanach:2012vj}). This is in general a challenging subject, and even R-parity conserving signatures often demand non-standard approaches~\cite{Plehn:2010st,Plehn:2011tf,Bai:2012gs,Han:2012fw,Kaplan:2012gd}, because regular jets+MET searches simply fail. The subject of RPV natural SUSY has received little attention thus far, and its collider signatures are still largely unexplored (see however searches by CMS where very little or no MET in the signal region is required~\cite{CMS:ssdilep,Chatrchyan:2012sa}). It would be interesting to study more signatures characterizing natural SUSY with baryon-number violation or lepton-number violation, as well as the more challenging spectrum orderings for the system introduced in this paper (lightest superpartner being sbottom or Higgsino).

We are very hopeful that searches along the lines described above will soon be performed at the LHC, and will help further our understanding of the grand hypothesis of Naturalness.

\acknowledgments {The authors are grateful to Kaustubh Agashe, Raffaele D'Agnolio, Roberto Franceschini, David E. Kaplan, David Krohn, Sanjay Padhi, Natasha Panikashvili, Salvatore Rappoccio, Matt Reece, and Matt Schwartz for useful discussions. We are also grateful to Matt Reece for useful comments on the manuscript and to Graham Kribs who pointed out an error in a previous version of this paper. CB and RS are grateful to support from the National Science Foundation under grant PHY-0910467 and by the Maryland Center for Fundamental Physics. AK is supported by the National Science Foundation under grant PHY-0855591.}

\bibliography{lit}
\bibliographystyle{jhep}
\end{document}